\documentclass[doublecol]{customized_epl2} 

\usepackage{graphicx}
\usepackage{amsmath}
\usepackage{amssymb}
\usepackage{amsfonts}
\usepackage{braket}
\usepackage{dsfont} 	
\usepackage{nicefrac} 	
\usepackage{lineno}
\usepackage{placeins}
\usepackage{dblfloatfix}
\usepackage{adjustbox}
\usepackage{xcolor}
\usepackage{paralist} 	
\usepackage{lipsum}
\usepackage{verbatim}
\usepackage{hyperref}
\usepackage[normalem]{ulem}
\usepackage{mathtools}

\usepackage[activate={true,nocompatibility},final,tracking=true,kerning=true,spacing=true,factor=1100,stretch=10,shrink=25]{microtype}
\SetTracking{encoding={*}, shape=sc}{0}

\hypersetup{
    bookmarks=true,         
    bookmarksopen=true,			
    bookmarksopenlevel=1,		
    unicode=false,          
    pdftoolbar=true,        
    pdfmenubar=true,        
    pdffitwindow=false,     
    pdfstartview={FitH},    
    pdftitle={Review of the DRAG framework},    
    pdfauthor={Lukas S. Theis},     
    colorlinks=true,       
    linkcolor=blue,          
    citecolor=blue,        
    urlcolor=blue           
}

\definecolor{lukas}{rgb}{0,0,1}
\definecolor{fwm}{rgb}{0,1,0}
\definecolor{shai}{rgb}{0,0.7,0.5}
\definecolor{felix}{rgb}{0.5,0.7,1}

\newcommand{\commut}[2]{[ #1 , #2 ]} 			
\newcommand{\op}[1]{\hat{#1}}								
\newcommand{\id}{\op{\mathds{1}}} 							
\renewcommand{\cos}[1]{\mathrm{cos}\LR{#1}} 				
\newcommand{\expInline}[1]{\mathrm{exp}(#1)}				
\newcommand{\dg}{^\dagger}

\newcommand{\del}{\mathrm{\Delta}}						
\newcommand{\ketbra}[2]{\ket{#1}\!\bra{#2}}				

\newcommand{\LR}[1]{\left(#1\right)} 					
\newcommand{\LRs}[1]{\left[#1\right]} 					
\newcommand{\abs}[1]{\left\lvert #1 \right\rvert}		
\newcommand{\normInline}[1]{\lVert #1 \rVert}		
\newcommand{\TDn}[3]{\frac{\text{d}^{#3} #1}{\text{d} #2^{#3}}} 	
		
\newcommand{\reffig}[1]{Fig.\,\ref{fig:#1}}

\newcommand{\refref}[1]{Ref.\cite{#1}}

\title{Counteracting systems of diabaticities using DRAG controls: The status after 10 years}
\shorttitle{Reviewing the DRAG framework} 
\author{L. S. Theis\inst{1} \and F. Motzoi\inst{2}\thanks{e-mail: \email{felix.motzoi@phys.au.dk}} \and S. Machnes\inst{1} \and F. K. Wilhelm\inst{1}}
\shortauthor{L. S. Theis \etal}

\institute{
	\inst{1} Theoretical Physics, Saarland University, 66123 Saarbr{\"u}cken, Germany\\
	\inst{2} Department of Physics and Astronomy, Aarhus University, Aarhus, Denmark
}


\abstract{The task of controlling a quantum system under time and bandwidth limitations is made difficult by unwanted excitations of spectrally neighboring energy levels. In this article we review the Derivative Removal by Adiabatic Gate (DRAG) framework. DRAG is multi-transition variant of counterdiabatic driving, where multiple low-lying gapped states in an adiabatic evolution can be avoided simultaneously, greatly reducing operation times compared to the adiabatic limit. In its essence, the method corresponds to a convergent version of the superadiabatic expansion where multiple counterdiabaticity conditions can be met simultaneously. When transitions are strongly crowded, the system of equations can instead be favorably solved by an average Hamiltonian (Magnus) expansion, suggesting the use of additional sideband control. We give some examples of common systems where DRAG and variants thereof can be applied to improve performance.}

\begin{document}

\maketitle

\section{Introduction}
The Derivative Removal by Adiabatic Gate (DRAG) technique \cite{Motzoi2009,Gambetta2011,Motzoi2013} was developed in the context of the emerging technology of high-precision superconducting quantum devices.  With coherence times of these systems improving dramatically towards the end of the first decade in twenty-first century, it became a promising possibility to address desired quantum transitions in the systems with increasing spectral resolution \cite{Jirari2005,Rebentrost2009,Safaei2009,Khani2009}.  However, very fast pulses were needed which was a problem both in terms of microwave shaping technology in a highly cooled environment \cite{Motzoi2011} and in terms of the rich level structure of nonlinear superconducting quantum circuits, which involves unwanted coupling to so-called `leakage' energy levels \cite{Steffen2003,Chow2009}.  The basic DRAG idea was to augment a simple smooth Rabi pulse $\Omega(t) \op{\sigma}_x$ with an off-quadrature auxiliary pulse with a simple dependence $\propto \partial_t \Omega(t) \op{\sigma}_y /\del$, where $\del$ is the gap energy to the nearest excited state.

The basic mechanism behind the correction is the removal of diabatic errors so that the system couples to the leakage subspace only adiabatically, returning back to the computational (`qubit') space by the end of the pulse.  Such ideas have a rich history, including the first application to removing leakage from STIRAP pulses \cite{Unanyan1997}, to its generalization to a broader class of problems in \cite{Demirplak2003,Demirplak2008}, to the formulation in terms of transitionless dynamics in \cite{Berry2009}, and finally to a general categorization under the framework of `Shortcuts to Adiabaticity' (STA) \cite{Torrontegui2013}.

Although DRAG is closely related to these ideas, they are not interchangeable and functionally solve different kinds of problems in quantum mechanics:  (I)  The DRAG framework is a convergent expansion that allows removing series of errors that differently affect different portions of the Hilbert space and operators therein.  Thus, functionally, it is perhaps closest to the transitionless superadiabatic driving technique of \cite{Ibanez2013}, based on the superadiabatic expansion \cite{Garrido1964,Berry1987}.  (II) The expansion allows the solution of not just one STA but can remove a system of diabatic errors to a manifold of unwanted low-lying gap states. In this sense, it is a powerful extension of STA methodology. (III) While STA usually deals with problems of adiabatic passage techniques, DRAG is equally well applicable to resonant driving problems, also known as spectral selectivity problems, where one can think of an `adiabatic elimination' of fast subspaces while only perturbatively affecting the near-degenerate subspace where resonant driving occurs.

The method was first tested in \cite{Chow2010,Lucero2010}.  Since then it has become a standard tool in superconducting qubit experiments \cite{DiCarlo2009,Bianchetti2010,Ballester2012,Chow2012,Magesan2012,Fedorov2012,Strand2013,Corcoles2013,Walter2017,Chen2016}.  There have been numerous implementations, extensions, and applications to different physical systems in the intervening ten years since we first presented the ideas.  Other techniques to deal with harmful transitions were developed\cite{Forney2010,Economou2015} but do not feature the flexibility of DRAG pulses, particularly with regards to spectral selectivity. We review here some of the main developments, including: connections drawn to other STA methods \cite{Torrontegui2013}, a better understanding of the convergence properties of the expansion \cite{Motzoi2013}; the application to spins, optical lattices and Rydberg atoms \cite{Theis2016b,Motzoi2013,Zhuang2013,Yang2017}, and the development of the closely related technique of WAHWAH \cite{Schutjens2013,Theis2016}.

The article is structured as follows. We first introduce the reader to relevant and related control frameworks, in particular the adiabatic theorem, superadiabaticity and counterdiabaticity. In the subsequent section, we then present a general mathematical formulation of how DRAG solutions can be derived exactly, using iterated frames. In particular, we show how to deal with diabatic errors on multiple and/or uncontrolled transitions.  However, for typical multi-level problems, it can be intractable to exactly solve the equations obtained from this general theory. We therefore continue to present three useful expansions to obtain perturbative DRAG solutions. We end the article with a review  of significant, experimentally motivated examples to demonstrate the general applicability of DRAG and show how perturbative solutions are obtained for a given system. 

\section{Review of adiabatic control techniques\label{sec:techintro}}
Consider a Hamiltonian $\op{H}_0(t)$ with eigenvalues $E_n(t)$ and eigenstates $\ket{n_0(t)}$, respectively. The adiabatic theorem \cite{Nenciu1980,Kato1950,Born1928,Berry1987} states that if the system is initially prepared in some instantaneous eigenstate $\ket{n_0(t)}$, the state evolved according to Schr{\"o}dinger's equation $i\partial_t \ket{\psi(t)}  = \op{H}_0(t)\ket{\psi(t)}$ will follow this eigenstate, i.e. $\ket{\psi(t)}  =\ket{n_0(t)}$ up to a phase, provided the Hamiltonian $\op{H}_0(t)$ changes sufficiently slowly ($\hbar\equiv 1$). This is referred to as adiabatic evolution. We refer the reader to \refref{Albash2018} for a detailed and rigorous analysis of different formulations of the theorem. It is convenient to transform to the adiabatic frame of $\op{H}_0$, that is the diagonal basis $\{\ket{\phi_n}\}$, by a unitary transformation $\op{V}_0(t)=\sum_ne^{i\varphi_n(t)}\ketbra{n_0(t)}{\phi_n}$, according to
\begin{align} \label{eq:adiabatic_basis}
  \op{H}_{\rm eff}(t) & =  \op{V}\dg_0(t) \op{H}_0(t) \op{V}_0(t) + i\dot{\op{V}}\dg_0(t) \op{V}_0(t).
\end{align}
The \emph{inertial term} $\op{I}(t)\equiv i (\partial_t \op{V}\dg_0(t)) \op{V}_0(t)$ is the source of transitions between instantaneous eigenstates of $\op{H}_0$ (we refer to these as \emph{diabatic errors}). If the adiabatic theorem can be applied, we may neglect the inertial term $\op{I}(t)$ in Eq.\eqref{eq:adiabatic_basis} up to a geometric phase \cite{Berry1984}, canceled for convenience via the free phase $\varphi_n(t)=\int_0^t\braket{n(t') | \partial_{t'}n(t')}\upd t'$ . Time evolution of the diagonal Hamiltonian $\op{D}(t)\equiv\op{V}\dg_0(t) \op{H}_0(t) \op{V}_0(t)=E_n(t)\ketbra{\phi_n}{\phi_n}$ is then given straightforwardly by
\begin{align}\label{eq:propagator_diag}
    \op{U}(t) =\sum\limits_{n}e^{-i\int_0^t E_n(t')\upd t'}\ketbra{\phi_n}{\phi_n},
\end{align}
where the evolution in the original basis is $\op{V}_0(t)\op{U}(t)\op{V}\dg_0(0)$. 

\subsection{Superadiabaticity\label{sec:superadiabaticity}}
For practical applications of system control, Hamiltonians often do not change slowly enough to justify an adiabatic approximation as given by Eq.\eqref{eq:propagator_diag}, i.e. neglecting the diabatic term $\op{I}(t)$. In order to properly quantify the finite rate of change theory needs to be extended accordingly. To this end, Berry provided a useful framework under the name of superadiabaticity\cite{Berry1987,Deschamps2008}, formulated also 20 years earlier by Garrido\cite{Garrido1964} in the context of adiabatic invariants. The method amounts to finding corrections to the Hamiltonian that account for finite inertial terms, by utilizing a sequence of iterative adiabatic transformations. That is, analogously to Eq.\eqref{eq:adiabatic_basis}, we define the $j$-th adiabatic frame Hamiltonian recursively as
\begin{align}\label{eq:superadiabaticity_1}
	\op{H}_{j} & = \op{V}_{j-1}\dg \op{H}_{j-1} \op{V}_{j-1} + i\dot{\op{V}}_{j-1}\dg \op{V}_{j-1}, \quad j \geq 1,
\end{align}
We assume all operators are (implicitly) time-dependent from this point onward. Each $\op{V}_j$ diagonalizes the error term $\op{I}_{j-1} \equiv i(\partial_t\op{V}_{j-1}\dg) \op{V}_{j-1}$, producing a new diagonal Hamiltonian $\op{D}_{j} \equiv \op{V}_{j}\dg \op{H}_{j} \op{V}_{j}$ and new inertial term $\op{I}_{j}$.

Unfortunately, these iterative transformations eventually begin to diverge\cite{Deschamps2008} from the actual dynamics of $\op{H}_0$ and so the expansion must be truncated.  The accuracy of a given frame can be quantified by the adiabatic quality factor $Q_j \equiv \int_0^{t_g}\normInline{\op{D}_j}/\normInline{\op{I}_j}$.  The highest value of $Q_j$ corresponds to the optimal frame, after which the series starts to diverge. However, for the purposes of high-fidelity quantum control, this frame will often be insufficient to meet accuracy requirements and therefore motivated the development of more accurate control methods over the last twenty years.  Nonetheless, the insight from the superadiabatic expansion will be crucial to using the DRAG framework which relies on the existence of such equivalent-frame adiabatic transformations. 

\subsection{Counterdiabaticity\label{sec:counterdiab}}
Counterdiabaticity (also: transitionless quantum driving) \cite{Torrontegui2013,Berry2009} comprises a technique to construct a Hamiltonian which drives states that exactly follow a desired trajectory. There are many choices for this counterdiabatic Hamiltonian $\op{H}_{\rm cd}$, e.g. due to different phases in the exponent of Eq.\eqref{eq:propagator_diag}, but the most straightforward is to exactly cancel the diabatic error term $\op{I}(t)$ . To this end, we augment the Hamiltonian $\op{H}_0$ to become
\begin{align}
  \op{\tilde{H}}_{0}(t) & = \op{H}_0(t) + \op{H}_{\rm cd}(t),\nonumber\\
  \op{H}_{\rm cd}(t)&=-\op{I}(t)
\end{align}
which renders Eq.\eqref{eq:propagator_diag} exact.

\section{Removing systems of multiple inertial terms with DRAG}
Combining the superadiabatic series with counterdiabaticity gives rise to the ability to solve systems with many unwanted diabatic transitions, including those that are not contained in the control Hamiltonian. This basic idea underlies the DRAG framework\cite{Motzoi2009,Gambetta2011,Motzoi2013} and related methodology \cite{Ibanez2013}. The DRAG transformations can take the form of a single dressing transformation, or an iterative frame expansion like the superadiabatic one. 

\subsection{Basic iterative framework}
Using Eq.~\eqref{eq:superadiabaticity_1}, we successively diagonalize the initial Hamiltonian as before. The expansion is now truncated at the $N$-th frame, which we refer to as the \emph{DRAG frame}, and rendered exact by the addition of a driving term which cancels $\op{I}_N=i(\partial_t\op{V}_{N}\dg)\op{V}_{N}$. Thus, transforming back to the initial Schr{\"o}dinger picture, i.e.~the frame of $\op{H}_0$ (the \emph{lab frame}), the respective counterdiabatic correction is 
\begin{align}\label{eq:CD_Ham_iterative}
	\op{H}_{\rm cd} & = -i\op{W}_N\op{I}_N\op{W}_N\dg = -i\op{W}_N \dot{\op{V}}_N\dg \op{W}_{N-1}\dg
\end{align}
where $\op{W}_j\equiv\prod_{m=0}^{j} \op{V}_m$ gives the transformation from first to $j$-th frame. A necessary condition for Eqs.~\eqref{eq:superadiabaticity_1} and \eqref{eq:CD_Ham_iterative} to match the intended dynamics is that the DRAG frame must coincide with the lab frame at initial and final time, i.e. $\op{W}_N(0)=\op{W}_N(t_g)=\id$. In this frame, Eq.~\eqref{eq:CD_Ham_iterative} ensures that the effective Hamiltonian is leakage-less with respect to unwanted couplings, i.e.~$\braket{m|\op{\tilde{H}}_{N}|n}=0$ for all relevant states $\ket{m}\neq\ket{n}$. One should note, however, that the DRAG frame is a dressed frame, and only the DRAG frame is transitionless throughout; all others exhibit leakage during intermediate times.  
Nonetheless, if the DRAG frame is equivalent to the lab frame at the boundaries of the time window, this ensures that no population remains in the unwanted states of the lab frame as well. This condition can typically be met if the control fields smoothly vanish for $t\in\{0,t_g\}$.

Next, we apply the DRAG methodology to two important cases of quantum control:  First, when the diabatic error $\op{I}_N$, and hence the desired correction, do not match the operator form of a physical control in the lab. Second, when two different diabatic error terms ($\op{I}_N=\op{I}_N^{(1)}\oplus \op{I}_N^{(2)}$) need to be corrected by operator terms sharing a single time dependence, e.g. a single laser field. We will refer to these situations as \emph{uncontrolled} vs. \emph{overconstrained} transitions, respectively. Though these two problems are in fact operationally quite different, most systems will exhibit both kinds of diabaticities.

\subsection{Uncontrolled diabaticities\label{sec:expconstr}}
In order to find solutions that are confined to attainable controls of the system we decompose the correction Hamiltonian \eqref{eq:CD_Ham_iterative} in terms of some available set of $k=1,\ldots,M$ non-overlapping controls. That is, we decompose the controllable Hamiltonian as $\op{H}^{\rm ctrl}(t)=\sum_ku_k(t)\op{h}_k+\mathrm{h.c.}$ with control fields $u_k$ and coupling terms $\op{h}_k=\ketbra{\psi_k^{(\rm to)}}{\psi_k^{(\rm from)}}$, so that
\begin{align}\label{eq:ctrl_set}
	\op{H}_{\rm cd}(t) & \equiv -i\sum_{k=1}^M \LRs{u_k(t)\op{h}_k+u_k^*(t)\op{h}_k\dg}.
\end{align}
We iteratively apply superadiabatic transformations to determine the control fields $u_k=\sum_ju_{k,j}$ in the counterdiabatic Hamiltonian \eqref{eq:ctrl_set}. The $u_{k,j}$ are contributions to the optimal control fields in the lab frame which cancel  diabatic errors from the $j$-th superadiabatic frame. To confine the corrections to the attainable controls, we calculate the respective overlap with the diabatic error $\op{I}_j$ for each successive superadiabatic iteration $j$, i.e.
\begin{align}\label{eq:DRAG1error}
  u_{k,j}(t) & = -i\Braket{\psi_k^{(\rm from)} | \op{W}_j\op{I}_j \op{W}_{j}\dg | \psi_k^{(\rm to)}},\\
  \op{I}_j^{\rm red} & = i\dot{\op{V}}_j\dg \op{V}_j - \sum_k  \op{W}_{j} \dg \LRs{u_{k,j}\op{h}_k+u_{k,j}^*\op{h}_k\dg} \op{W}_{j}\notag. 
\end{align}
The inertial term $\op{I}_j$ is now generally reduced with DRAG to $\op{I}_j^{\rm red}$ for $j>0$, i.e.~there is a partial cancellation if the counterdiabatic corrections do not map to attainable controls perfectly. Subsequent superadiabatic iterations are used to diagonalize $\op{I}_j^{\rm red}$ along the lines of \eqref{eq:superadiabaticity_1}, where $\op{I}_j^{\rm red}$ now replaces $i(\partial_t\op{V}_j\dg)\op{V}_j$. 
Whereas the standard sequence of superadiabatic transformations may often diverge (${Q_{\infty}}\rightarrow 0$), this series will (typically) converge to zero error as the diabatic error is iteratively reduced. For example, a 3-level system where one transition is driven and a second remains constant and uncontrolled  will exhibit this feature \cite{Petrosyan2017}. 

In practice, some transitions may share a common time-dependence; however, we omit this case here for clarity because it typically results in overconstrained transitions -- which will be treated in the next subsection. 

\subsection{Overconstrained diabaticities\label{sec:multileak}}
When error terms are not independently controlled (as is usually the case), this can lead to counterdiabatic expansions that do not converge, similarly to the superadiabatic series \eqref{eq:superadiabaticity_1}. This can be understood as a consequence of the fact that for long times $|\op{I}_{j+1}|\geq|\op{I}_{j}|$ \cite{Motzoi2013,Deschamps2008}, and so we must proceed carefully. For clarity, we consider here the case where all transitions in the system are controlled with a single global field, i.e.~$u_k(t)=u(t)$. The direct recursive solution \eqref{eq:DRAG1error} will now be replaced with the simultaneous constraints
\begin{align}\label{eq:DRAGnerrors}
    &u(t)  = \sum_j v_{j}(t) \,\,\, \text{and}\,\,\, \Braket{m | {{\op I}^{\rm red}}_N | n}= 0,\quad \forall m\neq n,  \\
    &\text{where again}  \quad {{\op I}^{\rm red}}_j  = i\dot{\op{V}}_j\dg \op{V}_j - \sum_k\op{W}_{j}\dg \LRs{v_{j}\op{h}_k+v_{j}^*\op{h}_k\dg} \op{W}_{j},\nonumber
\end{align}
where the $u_{k,j}$ from the previous case \eqref{eq:DRAG1error} were replaced by $v_j$ to emphasize the independence of $k$. 
The system \eqref{eq:DRAGnerrors} becomes fully constrained when the total number of frames $N$ equals the number of unwanted transitions $M$. A larger frame number can also be used, for example if some controls are not attainable.  In general this must be solved for all times, though one can often (e.g.~the perturbative limit) solve for all times simultaneously through a single system of $M$ algebraic equations (see also  \cite{Motzoi2018} where it is exact).

\begin{figure*}
	\centering
    \includegraphics[width=\textwidth]{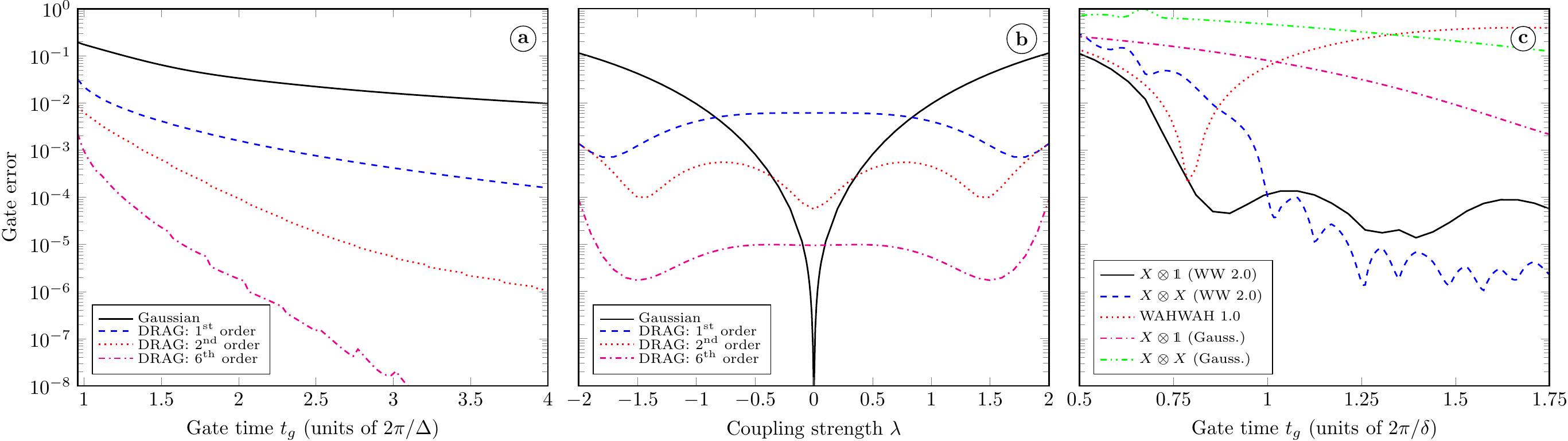}
    \caption{\label{fig:DRAG_perform}(a): Performance of unoptimized DRAG variants for removing leakage as a function of gate time derived from an iterative Schrieffer-Wolff expansion \eqref{eq:SW_single_leakage} to higher orders. Target : $\op{\sigma}_x$ rotation on qubit subspace. -- (b): Error for a $\op{\sigma}_x$ gate implemented in a gate time $t_g=4\pi/\del$ using DRAG pulses from (a) as a function of coupling strength $\lambda$ to the leakage level. -- (c): Gate error as a function of gate time for WAHWAH controls (1.0 and 2.0) for a two-qutrit system. Separate single-qubit rotations and simultaneous rotations both share a speed limit at about $2\pi/\delta$. }
\end{figure*}

\section{Perturbation methods with DRAG\label{sec:theory}}
The formulation of the counterdiabatic correction to the superadiabatic expansion is not generally analytically solvable, and for infinite dimensional systems even numerical solutions are intractable.  However, for durations much shorter than the adiabatic limit (but still longer than the inverse of the smallest energy gaps), it is still possible to obtain very high-fidelity solutions in the perturbative limit. We discuss three approximation methods, recalling that we restrict ourselves to non-overlapping, traceless controls for simplicity.

\subsection{Schrieffer-Wolff (SW) expansion\label{sec:problem}}
To expand the exact multi-state, DRAG solutions in \eqref{eq:CD_Ham_iterative}, \eqref{eq:DRAG1error} and \eqref{eq:DRAGnerrors} one can use a power series in the inverse gap energies, which we call $\del_j$. 
We express the diagonalization operator in the SW form \cite{Schrieffer1966},  $\op{W}(t)=\expInline{\op{S}(t)}$. Since the transition elements (the controls) are time-dependent, the SW transformation must be amended to include time-dependence \cite{Gambetta2011,Theis2017}. The effective Hamiltonian is then given by
\begin{align}\label{eq:SW_leakage}
	\op{H}_{\rm eff} & = \sum\limits_{n}\frac{1}{n!}\commut{\op{H}}{\op{S}}_n - i\sum\limits_{n}\frac{1}{(n+1)!}\commut{\dot{\op{S}}}{\op{S}}_n,
\end{align}
where $\commut{\op{A}}{\op{B}}_n=\commut{\commut{\op{A}}{\op{B}}_{n-1}}{\op{B}}$ and $\commut{\op{A}}{\op{B}}_0=\op{A}$. Contrary to the exact DRAG solutions in \eqref{eq:CD_Ham_iterative}, \eqref{eq:DRAG1error} and \eqref{eq:DRAGnerrors}, the SW method determines a \emph{single} frame transformation $\op{S}=\sum_j\op{S}^{(j)}$ to orders $j$ in the small parameter $u/\del$.  Let us start with the case of canceling a single harmful transition $\del_k$, corresponding to a term $u_{k}(t)\op{h}_k+ u_{k}^*(t)\op{h}_k\dg$ in the initial frame. The required generator of $\op{W}$ that diagonalizes the $j$-th order error is $\op{S}_k^{(j)}(t)=(w_{k,j}(t)\op{h}_k - w_{k,j}^*(t)\op{h}_k\dg)/\del_k$. The $w_{k,j}$ correspond to $j$-th order error terms in the dressed frame (so  $w_{k,0}=u_k$). In general, multiple transitions will be detrimental, and their total effect can be captured by $\op{S}^{(j)}=\sum_k\op{S}^{(j)}_k$.

To guarantee convergence one must be sure to count orders correctly\cite{Motzoi2013}. As we show in the next subsection $\mathcal{O}(\partial_t u_{k,j}(t))=\mathcal{O}(u_{k,j}(t)\del_k)$, so that the inertial term in \eqref{eq:SW_leakage} can be as important as the first term -- as is true for the superadiabatic expansion -- thus making counterdiabatic terms crucial.  Given this ordering of errors, one can determine counterdiabatic corrections $u_{k,j}$ to control $k$ from the dressed error $w_{k,j}$, either by direct application or by $j$-th order Taylor expansion of the inverse transformations.

In the spirit of the superadiabatic expansion one may alternatively cancel $k=1,\ldots,M$ transitions $\del_k$ using $MN$ generators $\op{S}_k^{(j)}$ iteratively with $j=1,\ldots,N$. The analog to the superadiabatic expansion is to pick instead $\op{W}=\prod \op{V}_l$ with $\op{V}_l=\expInline{\op{S}_k^{(j)}}$, where the indices $k$ and $j$ are uniquely combined into a single index $l\equiv (k,j)$ running from $1,\ldots,MN$.  
We compute the $l$-th effective as Hamiltonian recursively via the relation
\begin{align}\label{eq:SW_single_leakage}
	\op{H}_l & = \sum\limits_n \frac{1}{n!}\commut{\op{H}_{l-1}}{\op{S}_l}_n - i\dot{\op{S}}_l
\end{align}
and require that all $j$-th order terms be canceled before diagonalizing the next order in the small parameter. Indeed, for $j=1$, this corresponds to the first-order error expansion of the superadiabatic series, where transitions as before can also be uncontrolled or overconstrained. Thus, when a diabatic term cannot be canceled with a counterdiabatic Hamiltonian it remains the same order in the small parameter, but of one higher iteration order in the superadiabatic frame numbering. Depending on the available controls or chosen superadiabatic frame, the SW procedure may feature several solutions to remove a given order of error \cite{Gambetta2011}.  

\subsection{Fourier spectrum}
A common method in spectroscopy\cite{Warren1984} is to use the Fourier transform (FT) of the input fields to estimate the excitation of transition elements in the Hamiltonian.  Remarkably, this approach reproduces\cite{Motzoi2013} the first order ($j=1$) transitionless SW expansion, connecting the concepts of adiabaticity and spectroscopy. 

In the rotating frame of the energies in the system, transition elements will take the form $u_k(t)e^{-i\del_k t}\op{h}_k+u_k^*(t)e^{i\del_k t}\op{h}_k\dg$. The spectral response is given by the finite-time FT, that is $\mathcal{F}(u,\del) \equiv \int_0^{t_g}u(t)e^{-i\del t}\,\upd t$. The counterdiabaticity condition can be rephrased via the identity
\begin{align}\label{eq:spectrum-2}
    \mathcal{F}(u,\del) & = \mathcal{F}\LR{\frac{i^r}{\del^r}\TDn{u(t)}{t}{r},\del} = 0,\quad \forall r\geq 1,
\end{align}
which follows directly from $r$-fold integration by parts when all $r$ boundary terms go to zero.  Thus, the spectrum of a single control $u$ at frequency $\del_k$ will be canceled out when we add the counterdiabatic Hamiltonian $\op{H}_{\rm cd}(t)=-\frac{i}{\del_k}\frac{\mathrm{d}}{\mathrm{d}t}(u(t)\op{h}_k-\mathrm{h.c.})$, corresponding to $r=1$ in Eq.\eqref{eq:spectrum-2}, or any of the higher derivative corrections. Equality \eqref{eq:spectrum-2} also explains why both terms in \eqref{eq:SW_single_leakage} are of the same order. In particular, where the FT describes the dynamics well (small error, or $\mathcal{F}(u,\del)\ll 1$) is also where the effect of $|\op{S}_l \del|$ and $|\partial_t \op{S}_l |$ should be very similar.  This corresponds to the regime of high quality $Q$ for the optimal superadiabatic frame \eqref{eq:superadiabaticity_1}, where the divergence of the series can be explained by the additional harmful effect of higher order terms in SW beyond the FT approximation.

The $n$-th derivative in the FT solution \eqref{eq:spectrum-2} corresponds to the superadiabatic corrections from the $n$-th iteration \cite{Motzoi2018}, and, as in the cases above, we can combine these to solve for multiple unwanted diabatic terms simultaneously, with gaps $\del_k$ respectively. We choose the ansatz control field $u(t) = b(t) + \sum_{r=1}^{N}a_{r}i^r \frac{\mathrm{d}^r}{\mathrm{d}t^r}b(t)$, where $b(t)$ is any smooth base waveform such as a Gaussian pulse, to find a solution to the linear system of algebraic equations
\begin{align}\label{eq:DRAG_coeff_semiclassical}
	1 + \sum\limits_r a_{r}({\del_k})^{r} = 0\quad\quad \forall k,
\end{align}
and solve for the coefficients $a_r$. The base waveform $b(t)$ and its derivatives need to start and end at zero for Eq.~\eqref{eq:spectrum-2} to hold. Solving system \eqref{eq:DRAG_coeff_semiclassical} for $N$ transitions gives 
\begin{align}\label{eq:firstordersuperad}
	u = b-i\sum_k\frac{\partial_tb}{\del_k}-\sum_k\sum_{j\neq k}\frac{\partial_t^2b}{\del_k\del_j}+\ldots+\frac{(-i)^N\partial_t^Nb}{\del_1\del_2\cdots\del_N}.
\end{align}
Other solutions exist when derivatives of higher order than $N$ are used. Eq.~\eqref{eq:DRAG_coeff_semiclassical} also corresponds to the exact DRAG solution when driving systems of harmonic oscillators \cite{Motzoi2018}. 

\subsection{Magnus expansion\label{sec:wahwah}}
The FT forms the first order (average Hamiltonian) term in the so-called Magnus series\cite{Magnus1954,Blanes2009} which gives an exact analytic expression for the propagator over a finite time window, i.e. $\op{U}(t_g)=\expInline{\sum_0^{\infty} \op{\bar{H}}_j}$. The series has different convergence criteria than SW, because the integrals, rather than the time-instantaneous terms, need to be small. The Magnus series generally takes a more involved iterative form than SW, but often it is truncated at the second order, $\op{\bar{H}}_2\propto\int_0^{t_g}\int_0^{t_2}\commut{\op{H}(t_1)}{\op{H}(t_2)}\upd t_1\upd t_2$.  Note that the series does not intrinsically enforce adiabaticity, but counterdiabaticity can be built in via \eqref{eq:spectrum-2}.

In addition to counterdiabatic terms, unwanted transitions can be removed by solving the (underconstrained) diagonalization conditions obtained from the Magnus Hamiltonian terms using any extra controls (similarly to SW). In particular, any off-resonant error term can be directly removed by driving at the transition frequency with the opposite weight, i.e.~spectral shaping\cite{Economou2015,Warren1984}. A solution that combines counterdiabaticity with spectral shaping using the Magnus expansion is the Weak AnHarmonicity With Average Hamiltonian (WAHWAH) pulse sequence \cite{Schutjens2013,Theis2016}. To improve experimental practicality, it is often desirable to work with a small smooth basis of time-domain waveforms, e.g. derivatives of Gaussians or Fourier components \cite{Theis2016}.

\section{Physical Examples}
We review some experimentally relevant applications of the DRAG framework. The basic motivations and results are summarized for each. 

\subsection{Single-qubit leakage via Schrieffer-Wolff}
State-of-the-art superconducting qubits, such as transmon qubits, are well modeled by a standard nonlinear oscillator \cite{Koch2007, Khani2009}. Their $j$-th energy level in the rotating frame is given by $\omega_j(t)=j\delta(t)+\del_j$, with anharmonicities $\del_j$ and $\delta(t)=\omega_q(t)-\omega_d$ being the qubit detuning from the carrier. Typically, $\abs{\del_2} \sim 0.05\omega_q$, so that leakage to higher near-resonant states deteriorates performance. The rotating frame Hamiltonian can be written as\cite{Gambetta2011}
\begin{align}\begin{split}
	\op{H}(t) & = \sum\limits_{j=1}^{d-1}\LRs{\omega_j(t)\op{\Pi}_j + \sum\limits_{\alpha}^{x,y}\lambda_{j-1}\frac{u_{\alpha}(t)}{2}\op{\sigma}_{j-1,j}^{\alpha}}. 
\end{split}\end{align}
Here, we used the effective Pauli spin operators $\op{\sigma}_{j,k}^{x}=\ketbra{j}{k}+\ketbra{k}{j}$ and $\op{\sigma}_{j,k}^{y}=-i\ketbra{j}{k}+i\ketbra{k}{j}$ for $k>j$ and the projector $\op{\Pi}_j=\ketbra{j}{j}$. Utilizing expansion \eqref{eq:SW_leakage} we decouple the qubit subspace $\{\ket{0},\ket{1}\}$ from the remaining Hilbert space by choosing $\op{V}=\expInline{-i\op{S}}$ with
\begin{align}
	\op{S}(t) & = \sum_j s_{z,j}(t)\op{\Pi}_j + \sum\limits_{\alpha}^{x,y}\sum_{j<k} s_{\alpha,j,k}(t)\op{\sigma}_{j,k}^\alpha.
\end{align}
We expand each $s_{\alpha,j,k}(t)$ in a power series of a small parameter $\epsilon=1/\del_2t_g$ to obtain respective solutions $s_{\alpha,j,k}^{(n)}(t)$ to arbitrary order $n$. Following the calculations in \refref{Gambetta2011} we find conditional equations for the $s_{\alpha,j,k}^{(n)}(t)$ to any order. Note that these equations reveal a set of free parameters, allowing for multiple solutions to the same order in $\epsilon$. For instance, a prominent solution in lowest order features a $y$-only correction, that is $u_y=-\dot{u}_x\lambda_1/2\del_2$ and $\delta=0$. 

To obtain higher order solutions it is preferable to use iterative transformations along the lines of Eq.\eqref{eq:SW_single_leakage}. Their performance is depicted in \reffig{DRAG_perform}a as a function of pulse length. How the quality of solutions depends on the coupling strength $\lambda$ is illustrated in \reffig{DRAG_perform}b for $t_g=4\pi/\del$. Higher order solutions are taken from \cite{Motzoi2013}. Note also that when the 0-2 transition is controlled via an additional corresponding frequency component, the three-level system can be solved exactly (cf.~chapter 8 in \cite{Motzoi2012}).

\subsection{Crosstalk in multi-qubit and qutrit systems\label{sec:wahwah}} A standard quantum control problem is the selective addressability of a two level system in the spectral vicinity of other such systems, as occurs e.g. in nuclear magnetic resonance (NMR) \cite{Criger2012}.  This problem is well adapted to a multi-diabatic control solution where weighted sums of different derivatives cancel all excitations in aggregate on the unwanted transitions \cite{Motzoi2013}, as in solutions \eqref{eq:DRAGnerrors} or more simply \eqref{eq:firstordersuperad}. 

When additional, crowded leakage levels are present, it has been shown to be advantageous to use sideband modulated controls, e.g.~based on the Magnus expansion methodology.  The so-called WAHWAH solution \cite{Schutjens2013,Theis2016} uses a sideband modulated Gaussian principle control in conjunction with an auxiliary DRAG pulse. Typical parameters of such a scenario are qubit frequencies $\omega_{q,j}\sim 5\mathrm{GHz}$, qubit anharmonicities $\abs{\del} \sim 300\mathrm{MHz}$ and crowded transitions $\omega_{q,2} + \del =\omega_{q,1}+\delta$ with $\delta \sim \omega_q/100$. The $x$-quadrature of the control field is supported by first order DRAG $u_y = -\dot{u}_x/2\del$ to minimize leakage within a qubit, and sideband modulated to cancel crowded transitions, i.e.
\begin{align}\label{eq:wahwah_ctrl}
	u_x(t) & = A_{0}e^{-\frac{(t-t_g/2)^2}{2\sigma^2}}\left\{1-\cos{\omega_x \left[t-\frac{t_g}{2}\right]}\right\}.
\end{align}
Here, $A_0$ enforces the desired rotation angle and $\sigma$ gives the standard deviation. At first \cite{Schutjens2013}, a modulation with $\omega_x=\delta/2$ was suggested (we refer to this as WAHWAH 1.0). As shown in \reffig{DRAG_perform}c, these controls can achieve errors $\mathcal{O}(10^{-4})$, below conventional error thresholds.  However, WAHWAH 1.0 is limited to work for specific gate durations $t_g$, and only one qubit may be driven at once. To overcome these limitations, the method was generalized in \cite{Theis2016}, suggesting an optimal sideband modulation $\omega_x=\omega_x(t_g,\delta)$. The significant improvement over Gaussian controls is illustrated in \reffig{DRAG_perform}c for the original (1.0) and improved version (2.0) of WAHWAH pulses. For details, particularly about the implementation of simultaneous gates using a smooth basis of Sine functions, we refer the reader to \refref{Theis2016}. WAHWAH 1.0 was experimentally demonstrated in \cite{Vesterinen2014}.

\subsection{Experimental DRAG and Pulse Calibration\label{sec:OC}}
Turning to the experimental implementation \cite{Lucero2010,Chow2010} of DRAG pulses: In practice, actual system parameters differ somewhat from those assumed in theory due to characterization gaps, system drift, or unknown transfer functions affecting the input field shapes \cite{Motzoi2011}. As a simplification, we assume the low order terms in DRAG are easier to implement as their shape will be mostly maintained on entry into the dilution fridge.  Even so, many different low-order variants of DRAG have been found in the literature for third-level leakage \cite{Motzoi2009,Gambetta2011,Motzoi2013,Lucero2010}. This reduced functional form can further be optimized theoretically \cite{Theis2016} and/or through a closed-loop process experimentally \cite{Egger2014,Kelly2014}, to account for the effect of higher order terms and experimental uncertainties (or using more advanced gradient-free algorithms such as CMA-ES\cite{Hansen2003}).   A systematic experimental study of the tune-up of the prefactors in front of the functional forms for the control operators was performed in \cite{Chen2016}.

\begin{figure}
	\centering
  	\includegraphics[width=.95\linewidth]{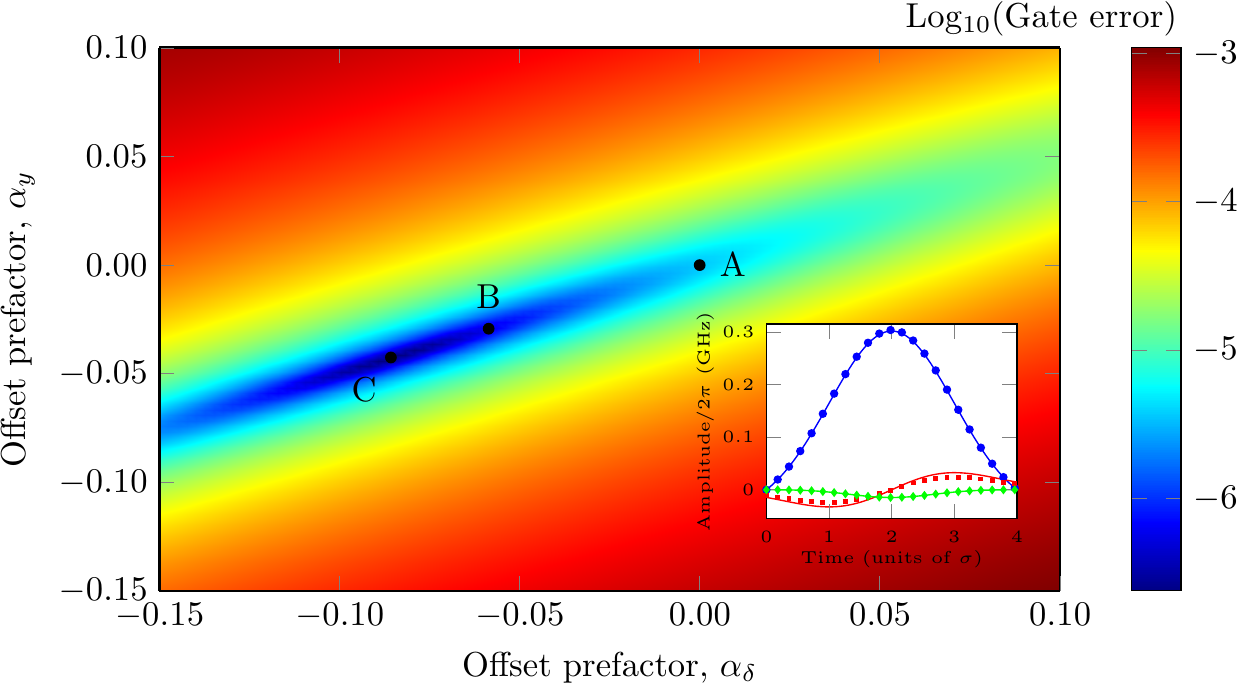}
  	\caption{\label{fig:calibration-landscape}A slice of the 3D calibration landscape for DRAG solution up to the first order in the small parameter to the qubit $\sigma_x$-gate leakage problem.  Point A and B denote \cite{Motzoi2009}'s and \cite{Gambetta2011}'s first-order solutions, respectively. Point C is the optimum for this control function subspace (here $\alpha_x=-0.0069$), with infidelity of $10^{-6.63}$. A successful calibration process will typically start at a known DRAG solution, i.e. points A or B, and conclude in point C. The inset illustrates the associated pulse shapes: markers represent the unoptimized shapes ($u_x$: {\color{blue} $\bullet$}, $u_y$: {\color{red} $\blacksquare$}, $\delta$: {\color{green} $\blacklozenge$}) whereas solid lines depict the corresponding optimal solution (C).}
\end{figure}

For instance, let us denote the Gaussian pulse implementing a $\sigma_x$ gate for the qubit by $G(t)$. Then the first order solution described in \cite{Motzoi2009,Gambetta2011,Motzoi2013} are parameterized by the limited functional basis $u_x \propto G$, $u_y\propto \partial_t G$ and $\delta\propto G^2$, which mimics the limited shaping control that can exist in experiment. None of the reported solutions are optimal within this functional basis. For some typical example parameters, infidelities may be further reduced from $10^{-5.28}$ to $10^{-6.63}$ by slightly adjusting the prefactors of the control fields. For example, \cite{Motzoi2009}'s first order DRAG solution may be transformed according to $u_x \longrightarrow (1+\alpha_x) u_x$ and similarly for $u_y$ and $\delta$, and then the constants $\alpha_x$, $\alpha_y$ and $\alpha_{\delta}$ are optimized. A discussion for why optimization within a severely restricted functional subspace may often be sufficient is given in \cite{Caneva2011} and follow-up publications. A schematic of the the optimization task involved in the calibration, as well as the shape of the associated controls, is shown in \reffig{calibration-landscape}.

\subsection{Interacting spectrally-crowded Rydberg atoms}
While DRAG has its origin in the field of superconducting qubits, it has also shown  promise in atomic systems (specifically trapped Rydberg atoms): Using a combination of DRAG control shaping and analytical optimization of the trap design, a proposal for errors below the conventional error threshold of $10^{-4}$ in a room temperature environment for entangling gates based on the Rydberg-blockade was made in\cite{Theis2016b}. Spectral shaping of the control field in form of Eq.\eqref{eq:firstordersuperad} is used to reduce leakage into several nearby Rydberg states other than the target state, and to further minimize blockade leakage. Such shaping techniques can also reduce gate times and leakage errors in other Rydberg-level based proposals for entanglement \cite{Petrosyan2017}.

\subsection{Motional states of atoms}
The anharmonic oscillator states that describe Josephson junction qubits are isomorphic to vibrational states of atoms in optical lattice potentials \cite{Khani2012}. Thus, most of the techniques in the above examples directly translate to non-harmonic traps \cite{Zhuang2013}.

\subsection{Microwave-entangled transmons}  Because coupled superconducting quantum systems retain their dense level structures, the ideas from the DRAG counterdiabatic framework are applicable to entangling operations as well (cf. chapters 9-10 in \cite{Motzoi2012}). This includes adiabatic passage type gates, used often with Xmon qubits \cite{Martinis2014}, but it is especially applicable to microwave entangling gates \cite{Ghosh2011,Ghosh2013,Economou2015}, which are transition-selective, e.g. the cross-resonance gate \cite{Kirchhoff2018}.

\subsection{Rotating wave approximation (RWA) correction} The ubiquitous approximation is used when resonantly driving quantum transitions \cite{Motzoi2011}. Physically, it corresponds to neglecting diabatic multi-photon transitions (e.g.~in Floquet theory \cite{Forney2010}) and these errors can also be suppressed \cite{Motzoi2013}. For `counter-rotating' terms at frequency $2\omega$, the DRAG solution is to correct Rabi driving with $\Omega(t)\op\sigma_x - \dot \Omega(t)\op\sigma_y/2\omega $.

\subsection{Spin resonance systems} The above treatments can also apply to two-level systems, for example for suppressing crosstalk (e.g. from global magnetic fields in NMR) or RWA \cite{Motzoi2013}. Application to quantum dots is given in \cite{Yang2017}.

\subsection{Applications to Lambda and sideband transitions}Multi-photon transitions are also used in adiabatic and resonant pulses, and can also suffer from diabatic errors to unwanted states \cite{Strand2013}.  This has inspired DRAG-like extensions to STIRAP pulses \cite{Baksic2016} as well as similar solutions to multi-photon Raman pulses (cf. chapter 10 in \cite{Motzoi2012}).

\subsection{Fast dispersive measurement} 
DRAG pulses have also found use for open-system control, being valuable in measurement dynamics where the readout apparatus must be used and reset quickly to avoid relaxation \cite{Bianchetti2010,Walter2017}.  In particular, Eq.~\eqref{eq:firstordersuperad} gives a solution to measuring $N$ qubit states via corresponding readout resonator modes where we replace the gap energies by inverse Lorentzians, $\Delta_j\equiv \delta_j + i\kappa/2$  \cite{Motzoi2018}. This forces fast (super)adiabatic following on a network of one or more decaying cavities, exactly solving Eq.~\eqref{eq:DRAGnerrors}.

\section{Summary}
The DRAG framework can be understood as an iterative counterdiabatic expansion which specifically allows to confine the solution space to attainable controls of a given system. DRAG solutions are constructed to simultaneously allow suppression of multiple transitions, also if available controls affect transitions other than intentionally driven ones. In general, exact analytic solutions to the underlying systems of equations are intractable.  We presented three common perturbative approaches to derive DRAG solutions of different orders: via an adiabatic Schrieffer-Wolff expansion, a spectral engineering approach and solutions derived from average Hamiltonian theory. Whereas the first two are based purely on expansions in terms of derivatives, the latter constitutes a different family of solutions (WAHWAH) which incorporates sideband modulations.  We note that different expansion methods treat higher orders differently: for instance, while the lowest order Magnus expansion reproduces the spectral engineering solutions that can be derived from a superadiabatic expansion, higher orders of both expansions differ significantly.

\section{Acknowledgements}
S. M. and F. K. W. acknowledge funding from the Intelligence Advanced Research Projects Activity (IARPA) through the LogiQ Grant No. W911NF-16-1-0114.

\bibliographystyle{eplbib.bst}
\bibliography{DRAG_review_EPL_etal.bib}

\end{document}